\documentclass[journal]{IEEEtran}
\usepackage{mathrsfs}
\usepackage{pifont}
\usepackage{bbding}
\usepackage{tipa}
\usepackage{dsfont}
\usepackage{algorithm}
\usepackage{algorithmic}
\usepackage{amsmath,epsfig}
\usepackage{stmaryrd}
\usepackage{amssymb}
\usepackage{amsfonts}
\usepackage{epic}
\usepackage{graphicx}
\ifCLASSINFOpdf
\else
\fi
\begin{document}\newcounter{fofo}
\newtheorem{theorem}{Theorem}
\newtheorem{proposition}{Proposition}
\newtheorem{definition}{Definition}
\newtheorem{lemma}{Lemma}
\newtheorem{corollary}{Corollary}
\newtheorem{remark}{Remark}
\newtheorem{construction}{Construction}

\newcommand{\supp}{\mathop{\rm supp}}
\newcommand{\sinc}{\mathop{\rm sinc}}
\newcommand{\spann}{\mathop{\rm span}}
\newcommand{\essinf}{\mathop{\rm ess\,inf}}
\newcommand{\esssup}{\mathop{\rm ess\,sup}}
\newcommand{\Lip}{\rm Lip}
\newcommand{\sign}{\mathop{\rm sign}}
\newcommand{\osc}{\mathop{\rm osc}}
\newcommand{\R}{{\mathbb{R}}}
\newcommand{\Z}{{\mathbb{Z}}}
\newcommand{\C}{{\mathbb{C}}}

%=======================================================title==============================================================================
%\title{Adaptive Subcarrier Pairing and Power Allocation with Limited Feedback for the OFDM Relay Networks}
\title{Adaptive Resource Allocation for Improved DF Aided Downlink Multi-user OFDM Systems}
%=======================================================author information=================================================================
%\author{\IEEEauthorblockN{Yong~Liu\IEEEauthorrefmark{1},
%Wen~Chen\IEEEauthorrefmark{1},
%Xiaopeng~Huang\IEEEauthorrefmark{2},
%%Montgomery Scott\IEEEauthorrefmark{3} and
%%Eldon Tyrell\IEEEauthorrefmark{4}}
%\IEEEauthorblockA{\IEEEauthorrefmark{1}Department of Electronic Engineering\\
%Shanghai Jiaotong University, Shanghai, 200240\\ Email: \{yongliu1982, wenchen\}@sjtu.edu.cn}.
%\IEEEauthorblockA{\IEEEauthorrefmark{2}Electrical and Computer engineering department\\Stevens Institute of Technology, New Jersey, USA\\
%Email: xhuang3@stevens.edu}}}
%Telephone: (800) 555--1212, Fax: (888) 555--1212}

\author{{Yong~Liu,}
        and~Wen~Chen,~\IEEEmembership{Senior member,~IEEE}
        % <-this % stops a space
\thanks{Manuscript received May 21, 2012; revised July 18, 2012, and accepted July 30, 2012. The associate
editor coordinating the review of this paper and approving it for
publication was Dusit Niyato.}
\thanks{Yong~Liu and Wen~Chen are
with the Department of Electronic Engineering, Shanghai Jiao Tong
University, Shanghai, 200240
PRC. e-mail: \{yongliu1982, wenchen\}@sjtu.edu.cn}
\thanks{This work is supported by national 973 project \#2012CB316106 and \#2009CB824904,
by NSF China \#60972031 and \#61161130529.}

% <-this % stops a space
%\thanks{This work is supported by NSF China \#60672067, by NSF Shanghai
%\#06ZR14041, by Shanghai-Canada NRC \#06SN07112, by Cultivation Fund
%of the Key Scientific and Technical Innovation Project, Ministry of
%Education of China \#706022, by Program for New Century Excellent
%Talents in University \#NCET-06-0386, and by PUJIANG Talents
%\#07PJ4046.}
}

\maketitle

%=======================================================abstract=================================================================
\begin{abstract}
%\boldmath
In this letter, we propose a joint resource allocation algorithm for
an OFDM-based multi-user system assisted by an improved
Decode-and-Forward (DF) relay. We aim at maximizing the sum rate of
the system by jointly optimizing subcarrier pairing, subcarrier
pair-user assignment, and power allocation in such a single DF relay
system. When the relay does not perform any transmission on some
subcarriers in the second phase, we further allow the source to
transmit new symbols on these inactive subcarriers. We effectively
solve the formulated mixed integer programming problem by using
continuous relaxation and dual minimization methods. Numerical
results verify the theoretical analysis, and illustrate the
remarkable gains resulted from the extra direct-link transmissions.
\end{abstract}


% IEEEtran.cls defaults to using math in the Abstract.


% \fi
\IEEEpeerreviewmaketitle

%=======================================================section1 introduction=======================================================
\section{Introduction}\label{sec:1}
\IEEEPARstart{P}{ower} allocation and subcarrier pairing have attracted much research attention in
OFDM-based relaying systems~\cite{IEEEconf:1}-\cite{IEEEconf:11}, due
to the limited budget of power, and independent fading on subcarriers in each hop.
Most of the previous works focus on single-user scenarios. Only a few literatures consider about multi-user cases~\cite{IEEEconf:6}-\cite{IEEEconf:9}. Authors in~\cite{IEEEconf:6}
investigate the problems of subcarrier allocation, subcarrier pairing and power allocation
in a relay aided two-hop uplink multi-user transmission model. Authors in~\cite{IEEEconf:7} work on the similar problem but for multi-relay channels. Authors in~\cite{IEEEconf:8} also study the resource allocation for the multi-user multi-relay model, and the optimal subcarrier allocation with QoS constraint is solved by using a graph theoretical approach. On the other hand, the joint resource allocation for relay aided downlink multi-destination networks is considered in~\cite{IEEEconf:9}-\cite{IEEEconf:11}.
However, in these works, the source always keeps silent
during the second phase, irrespective to whether the relay subcarriers
are active or not. In this paper, we consider an \emph{improved} DF relaying scheme.
For the circumstances where the relay does not
forward on some subcarriers because doing so does not improve
the sum rate, we further allow the source to transmit
extra symbols on these idle subcarriers in the source-destination link in the second
phase.
Authors in~\cite{IEEEconf:60} and~\cite{IEEEconf:80} consider a similar transmission scheme, but
for single-user scenario.
Jointly
optimizing subcarrier pairing, subcarrier pair-user assignment and power allocation for multi-user channels makes the problem more
general and difficult.
%
%
%
%
%Resource allocation using feedback can significantly improve performance of relay channels~\cite{IEEEconf:28},
%~\cite{IEEEconf:6}-\cite{IEEEconf:14}. Ahmed \emph{et al.} propose power
%control to minimize the outage probability of the AF in~\cite{IEEEconf:6}. Authors in~\cite{IEEEconf:30} present a Lloyd algorithm based codebook construction for DF system.

%=======================================================section2=====================================================================
\section{System Description}\label{sec:2}

The \emph{improved} OFDM DF diversity model is shown in Fig.~1, in which a source
communicates with $K$ users assisted by a single relay with \emph{N} subcarriers.
\begin{figure}[!t]
\centering
\includegraphics[width=3in,angle=0]{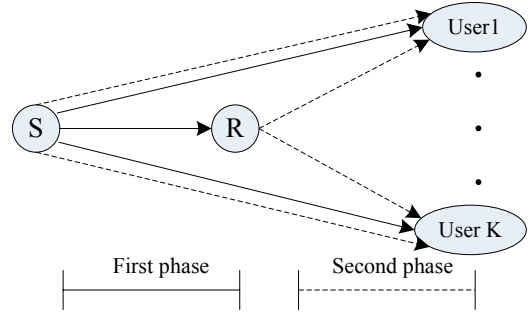}
\centering \caption{OFDM based improved DF diversity downlink
multi-user model.}\label{fig1}
\end{figure}
The channel gains are assumed
to be constant during two phases. In the first phase, the source broadcasts its signal while the relay and the destination listen. In the second phase, the relay transmits an encoded version of the received signal on some subcarriers, while the source transmits
a new modulated symbol on the subcarriers that are not used by the relay.
We use $\mbox{SP}(k,mn)$ to denote the subcarrier $m$ in the
first phase pairing with the subcarrier $n$ in the second phase, and the pair of channels $(m,n)$ is assigned to user $k$. We respectively denote by $h^{m}_{SD_{k}}$, $h^{m}_{SR}$ and $h^{n}_{RD_{k}}$ the channel coefficients of source-user $k$, source-relay, and relay-user $k$ on the corresponding subcarriers. If the relay keeps silent on subcarrier $n$ during the second phase, the channel coefficient of the extra source-destination link is denoted by $h^{n}_{SD_{k}}$. Then the corresponding normalized channel gains are respectively $\gamma^{m}_{S,D_{k}}=|h^{m}_{SD_{k}}|^{2}/\sigma_{k}^{2}$, $\gamma^{m}_{S,R}=|h^{m}_{SR}|^{2}/\sigma_{r}^{2}$,
$\gamma^{n}_{R,D_{k}}=|h^{n}_{RD_{k}}|^{2}/\sigma_{k}^{2}$ and $\gamma^{n}_{S,D_{k}}=|h^{n}_{SD_{k}}|^{2}/\sigma_{k}^{2}$, in which $\sigma^{2}_{k}$ and $\sigma^{2}_{r}$ are noise powers at the user $k$ and the relay.

Given a potential subcarrier user pair $\mbox{SP}(k,mn)$ in the
\emph{improved} DF diversity model, its achievable rate is given
in~(1), where
\begin{figure*}[!t]
\begin{displaymath} \tag{1}
R^{k,mn}=\begin{cases}
\;\frac{1}{2}\log_{2}\left(1+\gamma^{m}_{S,D_{k}}P^{m}_{S,D_{k}}\right)
+\frac{1}{2}\log_{2}\left(1+\gamma^{n}_{S,D_{k}}P^{n}_{S,D_{k}}\right), & \mbox{idle\, mode},\\
\;\frac{1}{2}\min\{\log_{2}\left(1+P^{m}_{S}\gamma^{m}_{S,D_{k}}
+P^{n}_{R,D_{k}}\gamma^{n}_{R,D_{k}}\right),\;\log_{2}\left(1+P^{m}_{S}\gamma^{m}_{SR}\right)\},&
\mbox{relaying \,mode}.
\end{cases}
\end{displaymath}
\rule{\linewidth}{0.5pt}
\end{figure*}
$P^{m}_{S,D_{k}}$ and $P^{n}_{S,D_{k}}$ denote the powers used by the
direct-link of $\mbox{SP}(k,mn)$ in the first and second phases, respectively. $P^{m}_{S}$ and $P^{n}_{R,D_{k}}$ respectively denote the source and relay powers in the relaying mode.

\setcounter{equation}{1}

Define a subcarrier pairing parameter $t_{m,n}\in\{0,1\}$, whose
value is $1$ if subcarriers $m$ and $n$ are paired, and $0$
otherwise. Denote $\pi_{k,mn}$ as the user assignment parameter,
whose value is $1$ if subcarrier pair $(m,n)$ is assigned to user
$k$, and $0$ otherwise. Since the condition for using the relay
depends not only on the channel gains but also on the power
allocation, we further denote $\varphi_{k,mn}\in\{0,1\}$ to show
whether the relay is used, i.e., the relay is used for
$\mbox{SP}(k,mn)$ if $\varphi_{k,mn}=1$. Otherwise, it is not used.
We aim to maximize the sum rate by jointly optimizing subcarrier
pairing, subcarrier pair-user assignment, and power allocation under
total power constraint.
%\begin{equation}   \nonumber
%\begin{split}
%\underset{\{\mathbf{t},\boldsymbol{\pi},\boldsymbol{\varphi},\mathbf{P}\}}\max&
%\sum^{N}_{m=1}\sum^{N}_{n=1}\sum^{K}_{k=1}R^{k,mn}, \\
%s.t.\;\;\mathbf{C}1: &\sum^{N}_{m=1}\sum^{N}_{n=1}t_{m,n}P^{m}_{S}\leq P_{S},   \; \mathbf{C}3:\sum^{N}_{m=1}t_{m,n}=1, \forall n, \\
%\mathbf{C}2:&\sum^{N}_{m=1}\sum^{N}_{n=1}t_{m,n}P^{n}_{R}\leq P_{R}, \; \mathbf{C}4:\sum^{N}_{n=1}t_{m,n}=1, \forall m.\\
%%&\mathbf{C}5:t_{m,n}\in\{0,1\}, \forall m,n.
%\end{split}
%\end{equation}
%\begin{equation}
%\begin{split}
%\underset{\{\mathbf{t},\boldsymbol{\pi},\boldsymbol{\varphi},\mathbf{P}\}}\max
%\sum^{K}_{k=1}\sum^{N}_{m=1}\sum^{N}_{n=1}R^{k,mn}.
%\end{split}
%\end{equation}

%Denote $P^{m}_{S}$ and $P^{n}_{R}$ as the source power and the relay power
%in the two phases with regard to $\mbox{SP}(m,n)$. Let  The noise powers at the destination and
%relay are assumed to be  respectively. Similarly
%we denote $P^{n}_{S}$ and  as the transmit power and the channel coefficient of the extra
%source-destination link, when subcarrier $n$ is not used by the relay.

%=======================================================section3=====================================================================

\section{Joint Resource Allocation}\label{sec:3}

\subsection{Optimization Problem Formulation}
Given a potential subcarrier pair $\mbox{SP}(k,mn)$ working in the relaying mode,
the achievable rate in (1) is maximized when
\begin{equation} \nonumber
\log_{2}\left(1+P^{m}_{S}\gamma^{m}_{SR}\right)=\log_{2}\left(1+P^{m}_{S}\gamma^{m}_{S,D_{k}}
+P^{n}_{R,D_{k}}\gamma^{n}_{R,D_{k}}\right),
\end{equation}
that is,
\begin{equation}
P^{m}_{S}\gamma^{m}_{SR}=P^{m}_{S}\gamma^{m}_{S,D_{k}}
+P^{n}_{R,D_{k}}\gamma^{n}_{R,D_{k}}.
\end{equation}
Let $P^{k,mn}=P^{m}_{S}+P^{n}_{R,D_{k}}$ for $\mbox{SP}(k,mn)$, we can express the achievable rate of $\mbox{SP}(k,mn)$ working in the relaying mode as
\begin{equation}
R_{R}^{k,mn}=\frac{1}{2}\log_{2}\left(1+\gamma^{k,mn}P^{k,mn}\right),
\end{equation}
by letting
\begin{equation}
\left\{ {\begin{array}{lp{5mm}l}
   {P^{m}_{S}=\frac{\gamma^{n}_{R,D_{k}}}{\gamma^{n}_{R,D_{k}}+\gamma^{m}_{SR}-\gamma^{m}_{S,D_{k}}}
   P^{k,mn}}, \\
   {P^{n}_{R,D_{k}}=\frac{\gamma^{m}_{SR}-\gamma^{m}_{S,D_{k}}}
   {\gamma^{n}_{R,D_{k}}+\gamma^{m}_{SR}-\gamma^{m}_{S,D_{k}}}P^{k,mn}},\\
   \gamma^{k,mn}=\frac{\gamma^{m}_{SR}\gamma^{n}_{R,D_{k}}}
{\gamma^{n}_{R,D_{k}}+\gamma^{m}_{SR}-\gamma^{m}_{S,D_{k}}}.
\end{array} } \right.
\end{equation}
%and
%\begin{equation}
%\gamma^{k,mn}=\frac{\gamma^{m}_{SR}\gamma^{n}_{R,D_{k}}}
%{\gamma^{n}_{R,D_{k}}+\gamma^{m}_{SR}-\gamma^{m}_{S,D_{k}}}.
%\end{equation}
Then the total end-to-end sum rate can be expressed as
\begin{equation}
\begin{split}
\mathbf{R}=\sum^{K}_{k=1}\sum^{N}_{m=1}
\sum^{N}_{n=1}&\frac{t_{m,n}}{2}\pi_{k,mn}\bigg\{\varphi_{k,mn}\log_{2}
\left(1+\gamma^{k,mn}P^{k,mn}\right)\\
&\quad\:\:+(1-\varphi_{k,mn})\big[\log_{2}\left(1+\gamma^{m}_{S,D_{k}}P^{m}_{S,D_{k}}\right)\\
&\qquad\qquad\qquad+\log_{2}\left(1+\gamma^{n}_{S,D_{k}}P^{n}_{S,D_{k}}\right)\big]
\bigg\},
\end{split}
\end{equation}
and the sum rate optimization can be formulated as
\begin{equation}
\begin{split}
\underset{\{\mathbf{t},\boldsymbol{\pi},\boldsymbol{\varphi},\mathbf{P}\}}\max&\mathbf{R}\\
s.t.\;\;\mathbf{C}1:&\sum^{N}_{m=1}t_{m,n}=1, \forall n, \quad\quad
\mathbf{C}2:\sum^{N}_{n=1}t_{m,n}=1, \forall m,\\
\mathbf{C}3:&\sum^{N}_{m=1}\sum^{N}_{n=1}\sum^{K}_{k=1}t_{m,n}\pi_{k,mn}\varphi_{k,mn}P^{k,mn}
+\\
&\quad\quad t_{m,n}\pi_{k,mn}(1-\varphi_{k,mn})(P^{m}_{S,D_{k}}+P^{n}_{S,D_{k}})\leq P_{t}, \\
\mathbf{C}4:&\sum^{K}_{k=1}\pi_{k,mn}=1, \forall m,n,\;
\mathbf{C}5:P^{k,mn},P^{m}_{S,D_{k}},P^{n}_{S,D_{k}}\geq0,\\
\mathbf{C}6: &t_{m,n},\;\pi_{k,mn},\;\varphi_{k,mn}\in\{0,1\},
\end{split}
\end{equation}
where
$\mathbf{P}=\left(P^{k,mn},P^{m}_{S,D_{k}},P^{n}_{S,D_{k}}\right)
\in (\mathbf{R}^3)^{N\times N}$, $\mathbf{t}=\left(t_{m,n}\right)\in
\mathbf{R}^{N\times N}$, $\boldsymbol{\pi}$ and
$\boldsymbol{\varphi}$ are matrices with entries $\pi_{k,mn}$ and
$\varphi_{k,mn}$ respectively. $\mathbf{C}1$ and $\mathbf{C}2$
correspond to the pairing constraint that each subcarrier $m$ in the
first phase only pairs with one subcarrier $n$ in the second phase.
$\mathbf{C}3$ denotes the total power constraint, in which $P_{t}$
denotes the sum of transmit powers at all the nodes. $\mathbf{C}4$
guarantees that each subcarrier pair can only be assigned to one
user.

Since the sum rate optimization mentioned above is a mixed integer
programming problem, which is hard to solve, we relax the integer
constraint $\mathbf{C}6:
t_{m,n},\,\pi_{k,mn},\,\varphi_{k,mn}\in\{0,1\}$ as $\mathbf{C}7:
t_{m,n},\,\pi_{k,mn},\,\varphi_{k,mn}\in[0,1]$ as
in~\cite{IEEEconf:11},\cite{IEEEconf:63}. Then the rate optimization
problem can be expressed as~(\ref{top}) with a new power constraint
$\mathbf{C}8:
\sum^{N}_{m=1}\sum^{N}_{n=1}\sum^{K}_{k=1}(S^{k,mn}+S^{m}_{S,D_{k}}+S^{n}_{S,D_{k}})\leq
P_{t},$ where we introduce new variables
$S^{k,mn}=t_{m,n}\pi_{k,mn}\varphi_{k,mn}P^{k,mn}$,
$S^{m}_{S,D_{k}}=t_{m,n}\pi_{k,mn}(1-\varphi_{k,mn})P^{m}_{S,D_{k}}$
and
$S^{n}_{S,D_{k}}=t_{m,n}\pi_{k,mn}(1-\varphi_{k,mn})P^{n}_{S,D_{k}}$
to denote the actual powers consumed by the relaying mode and idle
mode, respectively.
\begin{figure*}[!t]
\begin{equation} \label{top}
\begin{split}
&\underset{\{\mathbf{t},\boldsymbol{\pi},\boldsymbol{\varphi},\mathbf{S}\}}\max\sum^{K}_{k=1}\sum^{N}_{m=1}
\sum^{N}_{n=1}\frac{t_{m,n}}{2}\pi_{k,mn}\bigg\{\varphi_{k,mn}\log_{2}
\left(1+\gamma^{k,mn}\frac{S^{k,mn}}{t_{m,n}\pi_{k,mn}\varphi_{k,mn}}\right)+(1-\varphi_{k,mn})\cdot\\
&\;\;\;\bigg[\log_{2}\left(1+\gamma^{m}_{S,D_{k}}
\frac{S^{m}_{S,D_{k}}}{t_{m,n}\pi_{k,mn}(1-\varphi_{k,mn})}\right)+\log_{2}
\left(1+\gamma^{n}_{S,D_{k}}\frac{S^{n}_{S,D_{k}}}
{t_{m,n}\pi_{k,mn}(1-\varphi_{k,mn})}\right)\bigg]
\bigg\},
\end{split}
\end{equation}
\rule{\linewidth}{0.5pt}
\end{figure*}
To guarantee $t_{m,n},\pi_{k,mn}$ and $\varphi_{k,mn}$ being
integer-valued, we solve the problem by the dual
method~\cite{IEEEconf:63}. Since the objective function
of~(\ref{top}) is composed of three concave functions, (\ref{top})
is thus a standard convex programming problem. Therefore, the
optimization problem satisfies the time-sharing condition that
guarantees zero duality gap~\cite{IEEEconf:63}.

\subsection{Optimal Solution in Dual Domain}

Construct the Lagrange function for~(\ref{top}) as
\begin{equation} \label{Lag}
\begin{split}
L(&\mathbf{t},\boldsymbol{\pi},\boldsymbol{\varphi},\mathbf{S},\lambda)=\mathbf{\tilde{R}}-\\
&\lambda\left(\sum^{N}_{m=1}\sum^{N}_{n=1}\sum^{K}_{k=1}(S^{k,mn}
+S^{m}_{S,D_{k}}+S^{n}_{S,D_{k}})-P_{t}\right),
\end{split}
\end{equation}
where $\mathbf{\tilde{R}}$ denotes the optimization objective in~(\ref{top}), and $\lambda$ is the dual variable corresponding to $\mathbf{C}8$. Then the dual objective function is
\begin{equation} \label{dualo}
g(\lambda)=\underset{\{\mathbf{S},\mathbf{t},\boldsymbol{\pi},\boldsymbol{\varphi}\}}\max
L(\mathbf{t},\boldsymbol{\pi},\boldsymbol{\varphi},\mathbf{S},\lambda),
s.t.~\mathbf{C}1,~\mathbf{C}2,~\mathbf{C}4,~\mathbf{C}7,
\end{equation}
and the dual problem is
\begin{equation}  \label{Rd1}
\underset{\{\lambda\}}\min
g(\lambda)\quad s.t.\;\;\lambda\geq0.
\end{equation}
A dual function is always optimized by first optimizing some
variables and then optimizing the others. We first take derivatives of~(\ref{Lag}) with respect to $S^{k,mn}$, $S^{m}_{S,D_{k}}$, and $S^{n}_{S,D_{k}}$ and obtain
\begin{equation}  \label{S1}
\begin{split}
&S_{*}^{k,mn}=t_{m,n}\pi_{k,mn}\varphi_{k,mn}\left[\frac{1}{2\lambda}
-\frac{1}{\gamma^{k,mn}}\right]^{+},\\
&S^{m*}_{S,D_{k}}=t_{m,n}\pi_{k,mn}(1-\varphi_{k,mn})\left[\frac{1}{2\lambda}
-\frac{1}{\gamma^{m}_{S,D_{k}}}\right]^{+},\\
&S^{n*}_{S,D_{k}}=t_{m,n}\pi_{k,mn}(1-\varphi_{k,mn})\left[\frac{1}{2\lambda}
-\frac{1}{\gamma^{n}_{S,D_{k}}}\right]^{+},
\end{split}
\end{equation}
in which $[x]^{+}= \max\{0,x\}$.

If we denote the rate contribution of $\mbox{SP}(k,mn)$ to the Lagrange function~(\ref{Lag}) in the relaying mode and idle mode as $R_{m,n}^{R}$ and $R_{m,n}^{I}$ respectively, then they can be obtained by substituting~(\ref{S1}) into~(\ref{Lag}), that is
\begin{equation}  \label{rrri}
\begin{split}
R_{m,n}^{R}=&\frac{1}{2}\log
\left(1+\gamma^{k,mn}\tilde{S}_{*}^{k,mn}\right)
-\lambda\tilde{S}_{*}^{k,mn},\\
R_{m,n}^{I}=&\frac{1}{2}\log\left\{\left(1+\gamma^{m}_{S,D_{k}}\tilde{S}^{m*}_{S,D_{k}}\right)
\left(1+\gamma^{n}_{S,D_{k}}\tilde{S}^{n*}_{S,D_{k}}\right)\right\}\\
&-\lambda(\tilde{S}^{m*}_{S,D_{k}}+\tilde{S}^{n*}_{S,D_{k}}),
\end{split}
\end{equation}
in which we denote $\tilde{S}_{*}^{k,mn}=\left[\frac{1}{2\lambda}
-\frac{1}{\gamma^{k,mn}}\right]^{+}$, $\tilde{S}^{m*}_{S,D_{k}}=\left[\frac{1}{2\lambda}
-\frac{1}{\gamma^{m}_{S,D_{k}}}\right]^{+}$ and $\tilde{S}^{n*}_{S,D_{k}}=\left[\frac{1}{2\lambda}
-\frac{1}{\gamma^{n}_{S,D_{k}}}\right]^{+}$ for briefness. Easily we obtain the optimal solution for $\varphi_{k,mn}$ as
\begin{equation}    \label{optv}
\varphi_{k,mn}^{*}=\left \{\begin{array} {c@{\;
\;}l} 1,\quad & \mbox{when} \quad R_{m,n}^{R}>R_{m,n}^{I},\\
0,\quad & \mbox{otherwise}.
\end{array}
\right.
\end{equation}
The optimal $\varphi^{*}_{k,mn}$ tells us whether it is better to use relay for
the subcarrier-user pair $\mbox{SP}(k,mn)$. Substitute~(\ref{S1}), ~(\ref{optv}) and~(\ref{rrri}) into~(\ref{Lag}), the original Lagrange function~(\ref{Lag}) can be simplified as
\begin{equation} \label{Lag2}
\begin{split}
L(\mathbf{t},\boldsymbol{\pi},\lambda)=t_{m,n}\pi_{k,mn}\Pi_{k,mn}+\lambda P_{t},
\end{split}
\end{equation}
where $\Pi_{k,mn}=\varphi_{k,mn}^{*}R_{m,n}^{R}+(1-\varphi_{k,mn}^{*})R_{m,n}^{I}$.
We can observe from~(\ref{Lag2}) that, to maximize $L(\mathbf{t},\boldsymbol{\pi},\lambda)$, the subcarrier pair $(m,n)$ should be assigned to the user with maximum value of $\Pi_{k,mn}$, that is,
\begin{equation}   \label{optpi}
\pi_{k,mn}^{*}=\left \{\begin{array} {c@{\; :
\;}l} 1 & k=\arg\underset{k\in[1,.,K]}\max \Pi_{k,mn},\\
0 & \mbox{otherwise}.
\end{array}
\right.
\end{equation}
If we denote $\Pi_{mn}=\Pi^{*}_{k,mn}=\underset{k\in[1,.,K]}\max
\Pi_{k,mn}$, the dual objective function~(\ref{dualo})  can be
further simplified as the following linear optimization problem
\begin{equation} \label{dualo2}
\begin{split}
g(\lambda)=&\underset{\{\mathbf{t}\}}\max\sum^{K}_{k=1}\sum^{N}_{m=1}
\sum^{N}_{n=1}\left\{t_{m,n}\Pi_{mn}+\lambda P_{t}\right\},\\
s.t.\quad &\mathbf{C}1,\; \mathbf{C}2, \;\mathbf{C}7,
%&P^{m,n}>0  \forall m,n,\\
%&t_{m,n}\geq0 \forall m,n,\\
%&\sum^{N}_{n=1}t_{m,n}=1 \forall m
\end{split}
\end{equation}
which is well known as the \emph{two-dimensional assignment
problem}~\cite{IEEEconf:13}. The Hungarian
Algorithm~\cite{IEEEconf:13} is an efficient algorithm to solve such assignment problem with complexity that is a polynomial function in $N$. Without loss of generality, we can express the
assignment result as
\begin{equation}   \label{optt}
t_{m,n}^{*}=\left \{\begin{array} {c@{\; :
\;}l} 1 & (m,n)=Hung(\Pi_{mn}),\\
0 & \mbox{otherwise}.
\end{array}
\right.
\end{equation}

Finally, minimizing the dual function~(\ref{Rd1}) is required according to the standard Lagrange dual method, which can be solved with the subgradient method~\cite{IEEEconf:63}, that is,
\begin{equation}   \label{optl}
\begin{split}
\lambda^{(i+1)}=&\lambda^{(i)}-a^{(i)}\big(P_{t}-\\
&\sum^{N}_{m=1}\sum^{N}_{n=1}\sum^{K}_{k=1}(S^{k,mn}+S^{m}_{S,D_{k}}+S^{n}_{S,D_{k}})\big),
\end{split}
\end{equation}
where $i$ is the iteration number, $a$ is step
size. With the updated $\lambda^{(i+1)}$ in each iteration, we can respectively obtain the new power allocation vectors, $\varphi_{k,mn}$ and $\pi_{k,mn}$ by~(\ref{S1}),~(\ref{optv}) and~(\ref{optpi}), and then update $t_{m,n}$ with the Hungarian
Algorithm. The whole procedure can be described
as in Algorithm~1.

\vspace{-0.1cm}\hspace{-\parindent}\rule{\linewidth}{1pt}\vspace{-0.1cm}
\vspace{-0.25cm}{\bf Algorithm 1} The Proposed Resource Allocation Algorithm\\
\rule{\linewidth}{0.75pt}

\hspace{-\parindent}{\underline{Step 1}}: Initialize
$\lambda$, $\max_{iter}$ and set $i=1$,

\hspace{-\parindent}{\underline{Step 2}}: If $(i < \max_{iter})$, let
$a^{(i)}=0.01/\sqrt{i}$,

\hspace{-\parindent}{\underline{Step 3}}: Compute the rate contributions $R_{m,n}^{R}$ and $R_{m,n}^{I}$ by\\
\indent\indent\indent\;(\ref{rrri}) with $\lambda=\lambda^{(i)}$, and corresponding channel gains,

\hspace{-\parindent}{\underline{Step 4}}: Compute $\varphi^{*(i)}_{k,mn}$ by~(\ref{optv}) using
 $R_{m,n}^{I}=R_{m,n}^{I(i)}$ and\\
\indent\indent\indent\;$R_{m,n}^{R}=R_{m,n}^{R(i)}$,

\hspace{-\parindent}{\underline{Step 5}}: Compute $\pi_{k,mn}^{*(i)}$ by~(\ref{optpi}) using $R_{m,n}^{I}=R_{m,n}^{I(i)}$,\\
\indent\indent\indent\;$R_{m,n}^{R}=R_{m,n}^{R(i)}$, $\varphi^{*}_{k,mn}=\varphi^{*(i)}_{k,mn}$ and $\lambda=\lambda^{(i)}$,

\hspace{-\parindent}{\underline{Step 6}}: Compute $t_{m,n}^{*}$ by~(\ref{optt}) using $R_{m,n}^{I}=R_{m,n}^{I(i)}$,\\
\indent\indent\indent\;$R_{m,n}^{R}=R_{m,n}^{R(i)}$, $\varphi^{*}_{k,mn}=\varphi^{*(i)}_{k,mn}$ and $\pi_{k,mn}^{*}=\pi_{k,mn}^{*(i)}$,

\hspace{-\parindent}{\underline{Step 7}}: Compute $S_{*}^{k,mn}$, $S^{m*}_{S,D_{k}}$ and $S^{n*}_{S,D_{k}}$ by~(\ref{S1}) using\\
\indent\indent\indent\;$t_{m,n}=t^{(i)}_{m,n}$, $\pi_{k,mn}^{*}=\pi_{k,mn}^{*(i)}$, $\varphi^{*}_{k,mn}=\varphi^{*(i)}_{k,mn}$ and\\
\indent\indent\indent\;$\lambda=\lambda^{(i)}$,

\hspace{-\parindent}{\underline{Step 8}}: Compute $\lambda^{(i+1)}$
by~(\ref{optl}) using $\lambda^{(i)}$, $a^{(i)}$,
\indent\indent\indent\;$S_{*}^{k,mn}=S_{*(i)}^{k,mn}$,
$S^{m*}_{S,D_{k}}=S^{m*(i)}_{S,D_{k}}$,
$S^{n*}_{S,D_{k}}=S^{n*(i)}_{S,D_{k}}$,

\hspace{-\parindent}{\underline{Step 9}}: If
 $\frac{|\lambda^{(i+1)}-\lambda^{(i)}|}{|\lambda^{(i+1)}|}<\varepsilon$, exit and
 output the optimal\\
\indent\indent\indent\;solutions, otherwise set $i=i+1$ and go to Step $2$.
\\
%\hfill$\blacksquare$
\rule{\linewidth}{.75pt}
The complexity of each iteration in the proposed algorithm
is almost dominated by the Hungarian
Algorithm, which is $O(N^{3})$~\cite{IEEEconf:13}. On the other hand, the complexity of exhaustive searching method is $O(N!K^{N}2^{N})$. Obviously, the complexity of the proposed method is acceptable.

%=======================================================section4=====================================================================
\section{Simulation Results}\label{sec:4}
We consider an OFDMA system with $K=4$, and each subcarriers channel
experiencing flat-fading. We further assume
$\sigma_{r}^2=\sigma_{d}^2$ for simplicity. The step size $a^{(i)}$
for the subgradient method is set to be $\frac{0.01}{\sqrt{i}}$. The
source and relay are assumed to be placed on a horizontal line, and
$d_{SR}=10$. The four users are uniformly placed on the right
semicircle centered at the relay, and with radius $d_{RD}=5$.

Fig.~2 is obtained by comparing the proposed algorithm with several
schemes: equal power allocation without subcarrier pairing (\emph{EP
w/o SP}), optimal power allocation without subcarrier pairing
(\emph{OPA w/o SP}), equal power allocation with subcarrier pairing
(\emph{EP with SP}) and the joint power allocation and subcarrier
pairing for conventional DF that does not use the idle subcarriers
(\emph{Conventional DF}). We find that the proposed algorithm as
well as the \emph{Conventional DF} always outperform the others,
and the performance gaps increase as the SNR increases, which
implies that the power allocation and subcarrier pairing indeed
provides a substantial improvement to the system performance. On the other
hand, the performance of \emph{improved} relay protocol is superior
to that of conventional DF, which results from the extra direct-link
transmission in the second phase. For $SNR=4,~10,~18$dB, the sum
rate improvement percentage is $7.44\%,~6.24\%~4.7\%$, respectively.
We also find that the iterative algorithm quickly converges (e.g., 9 iterations).
\begin{figure}[!t]
\centering
\includegraphics[width=2.95in,angle=0]{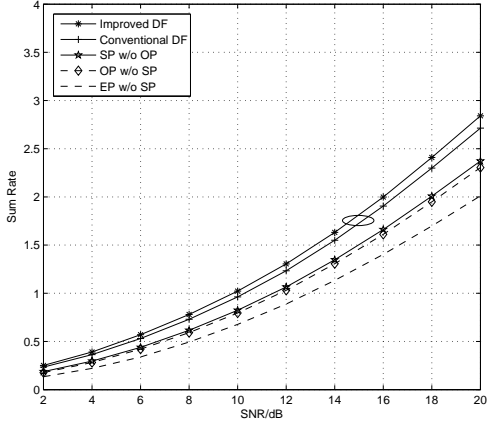}
\caption{Sum rate versus SNR for $N=4$, and $K=4$.} \label{fig2}
%the upper curve denotes dual problem solution, the lowest curve
%shows system rate without SP and optimal power allocation. The other
%four curves denote schemes $(ii)$, $(iii)$ and the proposed modified
%idle scheme with 2-bit feedback.
\end{figure}

%The extra rate contribution of the direct-link transmission in the second phase is showed in Fig.~3. We can find that, the larger extra power consumed is, the more rate contribution can be achieved.
%
%\begin{figure}[!t]
%\centering
%\includegraphics[width=3in,angle=0]{extra.eps}
%\caption{The extra rate contribution of the direct-link transmission in the second phase. $N=4,K=4$ is considered.} \label{fig2}
%%the upper curve denotes dual problem solution, the lowest curve
%%shows system rate without SP and optimal power allocation. The other
%%four curves denote schemes $(ii)$, $(iii)$ and the proposed modified
%%idle scheme with 2-bit feedback.
%\end{figure}

The sum rates of the OFDMA system versus $N$ for the conventional DF
and the \emph{improved} DF modes are illustrated in Fig.~3. For
$N=4,8,16,32,64$ respectively. $10000$ of such two-phase periods are
simulated. We observe that the rate gain of \emph{improved} DF
increases with $N$, which results from more frequency diversity and
pairing flexibility of large $N$.

%The performances of the \emph{improved} AF with different feedback bits are also revealed. We observe that only a few bits feedback can
%achieve most of the gain of the perfect CSI case. We also
%notice that further increase in the feedback bits brings degressive
%returns, which implies that the feedback bits or the codebook size is not necessarily too large.

\begin{figure}[!t]
\centering
\includegraphics[width=2.95in,angle=0]{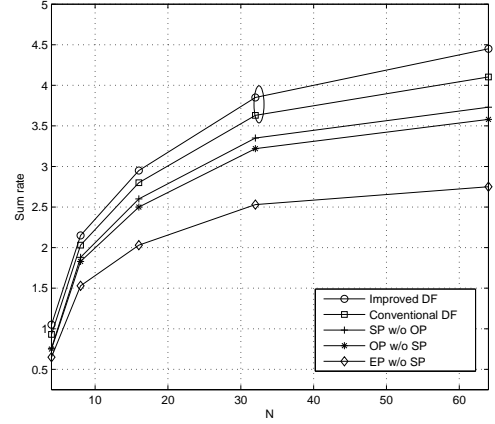}
\caption{Sum rate versus $N$ for $P_{t}=10$ and $K=4$.} \label{fig3}
\end{figure}

%\begin{figure}[!t]
%\centering
%\includegraphics[width=4in,angle=0]{lrate3.eps}
%\caption{Sum rate for
%$d=0.25,\,0.5,\,0.75$, with $2$-bit feedback.} \label{fig5}
%\end{figure}
%In order to exploit the sum rate for different relay
%locations, we simulate the rate with $d=0.25$,
%$d=0.5$ and $d=0.75$ respectively. Fig.~6 is achieved with $2$-bit feedback. We find that the \emph{improved} AF mode always outperforms others in any kind of $d$, and the channel condition of $R$-$D$ plays a more important role than the channel condition of $S$-$R$ in
%general.

%=======================================================section5=====================================================================
\section{Conclusion}\label{sec:5}

In this letter, we have proposed a joint resource allocation algorithm for
the OFDM \emph{improved} DF diversity channels, where the source is
allowed to transmit new messages on the idle subcarriers that are
not used in the second phase. With total power constraint, the joint
optimization has been formulated as a mixed integer programming problem.
We have solved this problem efficiently with polynomial complexity with the dual
method. Simulation results have verified that a remarkable rate gain can be
achieved with the extra direct-link transmissions.

%=======================================================section4=====================================================================

%\newpage

%\begin{figure}[!t]
%\centering
%\includegraphics[width=3in,angle=0]{rate2.eps}
%\caption{System sum rate versus SNR for the proposed improved AF
%relay scheme with different levels of feedback bits.
%%The upper curve
%%denotes the perfect CSI case, the lowest curve denotes scheme
%%without feedback, where the power is uniformly allocated. Others
%%curves demonstrate the effect of different feedback bits on sum
%%rate
%.} \label{fig2}
%\end{figure}

%\begin{figure}[!t]
%\centering
%\includegraphics[width=3in,angle=0]{rate04new.eps}
%\caption{The sum rate versus the number of subcarriers for the proposed schemes with fixed feedback bit of 2. $d=0.3$.} \label{fig4}
%\end{figure}

%\begin{figure}[!t]
%\centering
%\includegraphics[width=3in,angle=0]{rate06new.eps}
%\caption{The sum rate versus the relay location for the improved AF
%and conventional AF with $2$-bit feedback, where N=8.} \label{fig5}
%\end{figure}

\begin{thebibliography}{21}
\bibitem{IEEEconf:1}
M.~Herdin, ``A chunk based OFDM amplify-and-forward relaying
scheme for 4G mobile radio systems," in \emph{Proc.~IEEE Int.~Conf.~Communications
(ICC)}, Jun.~2006.


\bibitem{IEEEconf:2}
Y.~Cui, V.~Lau, and R.~Wang, ``Distributive subband allocation, power
and rate control for relay-assisted OFDMA cellular system with imperfect
system state knowledge," \emph{IEEE Trans. Wireless Commun.,} vol.~8, no.~10, pp.~5096-5102, Oct.~2009.


\bibitem{IEEEconf:3}
H.~Wan, W.~Chen, and J.~Ji, ``Efficient Linear Transmission Strategy for MIMO Relaying Broadcast Channels with Direct Links," \emph{IEEE Wireless Communications Letters}, vol.~1, no.~1, pp.~14-17, 2012.

%W.~Wang and R.~Wu, "Capacity Maximization for OFDM Two-Hop Relay System With Separate Power Constraints," \emph{IEEE Trans. Veh. Technol.}, vol.~58, no.~9, pp.~4943-4954, Nov.~2009.


%\bibitem{IEEEconf:4}
%S.~Senthuran, A.~Anpalagan, and O.~Das, "Cooperative Subcarrier and Power Allocation for a
%Two-Hop Decode-and-Forward OFCDM Based Relay Network," \emph{IEEE Trans. on Wireless
%Commun.}, vol.~8, no.~9, pp.~4797-4805, Sep.~2009.


\bibitem{IEEEconf:5}
Z.~Wang, W.~Chen, and J.~Li, ``Efficient Beamforming for MIMO Relaying Broadcast Channel with Imperfect Channel Estimation," \emph{IEEE Transactions on Vehicular Technology}, vol.~61, no.~1, pp.~419-426, 2012.


\bibitem{IEEEconf:70}
H.~Wan, and W.~Chen, ``Joint Source and Relay Design for Multi-user MIMO Non-regenerative Relay Networks with Direct Links," to be published in \emph{IEEE Transactions on Vehicular Technology}, 2012.
%Y.~Liu, W.~Chen, and X.~Huang, "Capacity Based Adaptive Power Allocation for the OFDM Relay Networks with Limited Feedback," \emph{IEEE International Conference on Communications (ICC2011)}, Jun.~2011.


\bibitem{IEEEconf:71}
G.~Sidhu, F.~Gao, W.~Chen, and W.~Wang, ``Joint Subcarrier Pairing and Power Loading in Relay Aided Cognitive Radio Networks," \emph{IEEE Wire.~Commu.~and Network.~Conf.~(WCNC2012)}, Apr.~2012.


\bibitem{IEEEconf:6}
G.~A.~Sidhu, F.~Gao, ``Resource Allocation for Relay Aided Uplink Multiuser OFDMA System," in Proc.~\emph{IEEE Conf.~on WCNC}, pp.~1-5, Apr.~2010.


\bibitem{IEEEconf:7}
H.~Jeong, J.~H.~Lee, H.~Seo, ``Resource Allocation for Uplink Multiuser OFDM Relay Networks with Fairness Constraints," in Proc.~\emph{IEEE Conf.~on VTC Spring}, pp.~1-5, Apr.~2009.

\bibitem{IEEEconf:8}
G.~Li, H.~Liu, ``Resource Allocation for OFDMA Relay Networks With Fairness Constraints," \emph{IEEE J.~Sel.~Areas Commun.}, vol.~24, no.~11, pp.~2061-2069, Nov.~2006.


\bibitem{IEEEconf:9}
B.~Fan, Wen.~Wang, Y.~Lin, L.~Huang, K.~Zheng, ``Subcarrier Allocation for OFDMA Relay Networks
with Proportional Fair Constraint," in Proc.~\emph{IEEE Conf.~on ICC}, pp.~1-5, 2009.

\bibitem{IEEEconf:10}
J.~Yuan, Q.~Wang, ``Adaptive Resource Allocation Schemes for Multiuser OFDMA
Nonregenerative Relay Networks," in Proc.~\emph{IEEE Conf.~on ICC}, pp.~1-5, 2010.

\bibitem{IEEEconf:11}
M.~Hajiaghayi, M.~Dong, and B.~Liang, ``Optimal channel assignment and power allocation for dual-hop multi-channel multi-user relaying," in Proc.~IEEE Conf.~on INFOCOM, Mini-Conference, pp.~76-80, Apr.~2011.

\bibitem{IEEEconf:60}
V.~Luc, L.~Jerome, O.~Onur, and Z.~Abdellatif, ``Rate-Optimized
Power Allocation for DF-Relayed OFDM Transmission under Sum and
Individual Power Constraints", \emph{EURASIP Journal on wireless
comm. and networking}, Vol.~2009, 2009.

\bibitem{IEEEconf:80}
H.~Boostanimehr, V.~K.~Bhargava, ``Selective Subcarrier Pairing and Power Allocation for DF OFDM Relay Systems with Perfect and Partial CSI," \emph{IEEE Trans. Wireless Commun.,} vol.~10, no.~12, pp.~4057-4067, Dec.~2011.



\bibitem{IEEEconf:63}
W.~Yu and R.~Lui, ``Dual methods for nonconvex spectrum optimization
of multi-carrier systems," \emph{IEEE Trans.~Commun.,} vol.~54, no.~7, pp.~1310-1322, Jul.~2006.


\bibitem{IEEEconf:13}
H.~W.~Kuhn, ``The hungarian method for the assignment problem," \emph{Naval Research Logistics}, vol.~52, pp.~7-21, 2005.

%\bibitem{IEEEconf:64}
%K.~Seong, M.~Mohseni, and J.~Cioffi, "Optimal resource allocation
%for OFDMA downlink systems," in \emph{Proc. IEEE Int.~Symp.~Inf.~Theory
%(ISIT)}, pp.~1394每1398, Jul.~2006.

%\bibitem{IEEEconf:3}
%R.~U.~Naber, H.~Bolcskei, F.~W.~Kneubuhler, "Fading relay channels:
%Performance limits and space-time signal design," \emph{IEEE J.
%Select. Areas Commun.}, vol.~22, no.~6, pp.~1099-1109, 2004.
%\bibitem{IEEEconf:2}
%A.~Host~Madsen and J.~Zhang, "Capacity bounds and power allocation
%in wireless relay channel," \emph{IEEE Trans. Inf. Theory}, vol.~51,
%pp.~2020-2040, Jun.~2005.

%\bibitem{IEEEconf:50}
%A.~Sendonaris, E.~Erkip, and B.~Aazhang, "User cooperation diversity-part I:
%system description," \emph{IEEE Trans.~Commun.}, vol.~51, no.~11, pp.~1927-1938, Nov.~2003.
%%\bibitem{IEEEconf:4}
%%C.~Hoymann, K.~Klagges, and M.~Schinnenberg, "Multihop communication
%%in Relay improved IEEE 802.16 networks," \emph{in Proc. IEEE
%%PIMRC'06}, pp.~1-4, Sep.~2006.
%
%%\bibitem{IEEEconf:22}
%%N.~R.~Van and P.~Ramjee, OFDM Wireless Multimedia Communications,
%%Boston: Artech House, 2000.
%
%%\bibitem{IEEEconf:5}
%%N.~Ahmed and B.~Aazhang, "Throughput Gains Using Rate and Power
%%Control in Cooperative Relay Networks," \emph{IEEE Trans. Commun.},
%%vol.~55, no.~5, pp.~656-660, April 2003.
%
%%
%%\bibitem{IEEEconf:23}
%%Y.~Liu, W.~Chen, "Power Allocation for the Fading Relay Channel with
%%Limited Feedback," in \emph{Proc. IEEE ICC＊2010}, May,~2010.
%
%
%
%
%\bibitem{IEEEconf:28}
%Y.~Liu, W.~Chen, X.~P.~Huang, "Capacity Based Adaptive Power
%Allocation for the OFDM Relay Networks with Limited Feedback," in
%\emph{Proc. IEEE ICC＊2011}, Jun.~2011.
%
%
%%\bibitem{IEEEconf:8}
%%I.~Hammerstrom and A.~Wittneben, "On the Optimal Power Allocation
%%for Nonregenerative OFDM Relay Links," in \emph{Proc. IEEE ICC＊06},
%%Jun.~2006.
%
%\bibitem{IEEEconf:9}
%I.~Hammerstrom and A.~Wittneben, "Power Allocation for Amplify and
%Forward MIMO-OFDM Relay Links," \emph{IEEE Trans. on Wireless
%Commun.}, vol.~6, no.~8, pp.~2798-2802, Aug. 2007.
%
%\bibitem{IEEEconf:25}
%M.~Herdin, "A chunk based OFDM amplify-and-forward relaying scheme
%for 4G mobile radio systems," in \emph{Proc. IEEE ICC 2006},
%Istanbul, Turkey, June 2006, vol.~10, pp.~4507-4512.
%
%\bibitem{IEEEconf:41}
%W.~Wang and R.~Wu, "Capacity Maximization for OFDM Two-Hop Relay System With Separate Power Constraints," \emph{IEEE Trans. Veh. Technol.}, vol.~58, no.~9, pp.~4943-4954, Nov.~2009.

%
%
%
%
%
%\bibitem{IEEEconf:10}
%W.~Ying, Q.~Xin-chun, and L.~Bao-ling, "Power Allocation and
%Subcarrier Pairing Algorithm for Regenerative OFDM Relay System," in
%\emph{Proc. IEEE Vehicular Technology Conf. (VTC)'07},
%pp.~2727-2731, April 2007.
%
%\bibitem{IEEEconf:26}
%Y.~Li, W.~Wang, J.~Kong, and M.~Peng, "Subcarrier pairing for
%amplify-and-forward and decode-and-forward OFDM relay links,"
%\emph{IEEE Commun. Lett.,} vol.~13, no.~4, pp.~209-211, APRIL 2009.
%
%%\bibitem{IEEEconf:27}
%%Y.~Li, W.~Wang, J.~Kong, W.~Hong, X.~Zhang, and M.~Peng, "Power
%%allocation and subcarrier pairing in OFDM-based relaying networks,"
%%in \emph{Pro. IEEE ICC 2008}, May 2008, pp.~2602-2606.
%
%\bibitem{IEEEconf:40}
%M.~Hajiaghayi, M.~Dong, and B.~Liang, "Optimal channel assignment and power allocation for dual-hop multi-channel multi-user relaying," in \emph{Proc.~IEEE Conf.~on INFOCOM, Mini-Conference}, pp.~76-80, Apr. 2011.
%
%
%
%
%
%
%
%
%\bibitem{IEEEconf:6}
%N.~Ahmed, M.~A.~Khojastepour, A.~Sabharwal and B.~Aazhang, "Outage
%Minimization With Limited Feedback for the fading Relay Channels,"
%\emph{IEEE Trans. Commun.}, vol.~54, no.~4, pp.~659-669, April~2006.
%
%%\bibitem{IEEEconf:7}
%%J.~Zhang, Q.~Zhang, C.~Shao, Y.~Wang, P.~Zhang, and Z.~Zhang,
%%"Adaptive optimal transmit power allocation for two-hop
%%non-regenerative wireless relay system," in \emph{Proc. IEEE
%%VTC＊04}, vol.~2, pp.~1213-1217, May~2004.
%
%
%%\bibitem{IEEEconf:11}
%%Y.~Ma, N.~Yi, and R.~Tafazolli, "Bit and Power Loading for OFDM
%%Based Three-Node Relaying Communications," \emph{IEEE Trans. on
%%Signal Processing}, vol.~56, pp.~3236-3247, July 2008.
%
%%\bibitem{IEEEconf:24}
%%Y.~Wang et al., "Power allocation and subcarrier pairing algorithm
%%for regenerative OFDM relay system," in \emph{Proc. IEEE VTC
%%2007-Spring}, Dublin, Ireland, Apr. 2007, pp.~2727每2731.
%
%
%%\bibitem{IEEEconf:42}
%%D.~Wang, Z.~ Li, and X.~Wang, "Jointly Optimal Subcarrier and Power Allocation
%%for Wireless Cooperative Networks over OFDM Fading Channels," to be published in \emph{IEEE Trans. Veh. Technol.}, 2011.
%
%
%\bibitem{IEEEconf:12}
%T.~T.~Kim and M.~Skoglund, "Diversity-multiplexing tradeoff in MIMO
%channels with partial CSIT," \emph{IEEE Trans. Inf. Theory},
%vol.~53, pp.~2743-2759, Aug. 2007.
%
%%\bibitem{IEEEconf:13}
%%T.~Yoo, A.~Goldsmith, "Capacity and Power Allocation for Fading MIMO
%%Channels With Channel Estimation Error," \emph{IRE Trans. Inform.
%%Theory}, vol.~52, no.~5, pp.~2203-2214, May~2006.
%
%\bibitem{IEEEconf:14}
%S.~Tatikonda, S.~Mitter, "The Capacity of Channels With Feedback,"
%\emph{IEEE Trans. Inform. Theory}, vol.~55, no.~1, pp.~323-349, Jan.~2009.
%
%\bibitem{IEEEconf:30}
%G.~Zhang, W.~Zhan, and J.~Qin, "Power Allocation in
%Decode-and-Forward Cooperative OFDM Systems Using Perfect and
%Limited Feedback", \emph{Chinese Journal of Electronics}, Vol.~19,
%No.~2, Apr.~2010.
%
%\bibitem{IEEEconf:43}
%W.~Yu and R.~Lui, "Dual methods for nonconvex spectrum optimization
%of multi-carrier systems," \emph{IEEE Trans.~Commun.,} vol.~54, no.~7, pp.~1310-1322, Jul.~2006.

%\bibitem{IEEEconf:32}
%L.~M.~C.~Hoo, B.~Halder, J.~Tellado, and J.~M.~Cioffi, "Multiuser
%Transmit Optimization for Multicarrier Broadcast Channels:
%Asymptotic FDMA Capacity Region and Algorithms," IEEE Trans.~
%Commun., vol.~52, no.~6, pp.~922-930, Jun.~2004.

%
%
%\bibitem{IEEEconf:18}
%J.~Choi and R.~W.~Heath, "Interpolation based transmit beamforming
%for MIMO-OFDM with limited feedback," \emph{IEEE Trans. Acoust.,
%Speech, Signal Processing}, vol.~53, no.~11, pp.~4125-4135, Nov.~2005.
%
%\bibitem{IEEEconf:19}
%D.~J.~Love and R.~W.~Heath, "OFDM power loading using limited
%feedback," \emph{IEEE Trans. Veh. Technol.}, vol.~54, no.~5,
%pp.~1773-1780, Sept.~2005.
%
%
%
%
%
%
%
%
%\bibitem{IEEEconf:15}
%C.~E.~Shannon, "Channels with side information at the transmitter,"
%\emph{IBM Journal Research and Dev.}, vol.~2, pp.~289-293, 1958.
%
%\bibitem{IEEEconf:16}
%S.~Bhashyam, A.~Sabharwal, and B.~Aazhang, "Feedback gain in
%multiple antenna systems," \emph{IEEE Trans. Commun.}, vol.~50,
%no.~5, pp.~795-798, May 2002.
%
%%\bibitem{IEEEconf:17}
%%S.~Tatikonda, S.~Mitter, "The Capacity of Channels With Feedback,"
%%\emph{IEEE Trans. Inform. Theory}, vol.~55, no.~1, pp.~323-349, Jan.
%%2009.
%
%
%
%%\bibitem{IEEEconf:29}
%%M.~Hajiaghayi, M.~Dong, and B.~Liang, "Using Limited Feedback in
%%Power Allocation Design for a Two-Hop Relay OFDM System", \emph{Pro.
%%IEEE ICC 2009}, pp.~1-6.
%
%
%
%\bibitem{IEEEconf:35}
%W.~Yu and J.~M.~Cioffi, "FDMA capacity of Gaussian multiple-access
%channels with ISI," IEEE Trans.~Commun., vol.~50, no.~1, pp.~102-111,
%Jan.~2002.
%
%
%
%
%\bibitem{IEEEconf:44}
%K.~Seong, M.~Mohseni, and J.~Cioffi, "Optimal resource allocation
%for OFDMA downlink systems," in \emph{Proc. IEEE Int.~Symp.~Inf.~Theory
%(ISIT)}, pp.~1394-1398, Jul.~2006.
%
%
%\bibitem{IEEEconf:31}
%V.~Luc, L.~Jerome, O.~Onur, and Z.~Abdellatif, "Rate-Optimized
%Power Allocation for DF-Relayed OFDM Transmission under Sum and
%Individual Power Constraints", \emph{EURASIP Journal on wireless
%comm. and networking}, Vol.~2009.
%%\bibitem{IEEEconf:24}
%%Y.~Wang et al., "Power allocation and subcarrier pairing algorithm
%%for regenerative OFDM relay system," in \emph{Proc. IEEE VTC
%%2007-Spring}, Dublin, Ireland, Apr. 2007, pp.~2727每2731.
%%
%%\bibitem{IEEEconf:25}
%%M.~Herdin, "A chunk based OFDM amplify-and-forward relaying scheme
%%for 4G mobile radio systems," in \emph{Proc. IEEE ICC 2006},
%%Istanbul, Turkey, June 2006, vol.~10, pp.~4507每4512.
%%
%%\bibitem{IEEEconf:26}
%%Y.~Li, W.~Wang, J.~Kong, and M.~Peng, "Subcarrier pairing for
%%amplify-and-forward and decode-and-forward OFDM relay links,"
%%\emph{IEEE Commun. Lett.,} vol.~13, no.~4, pp.~209-211, APRIL 2009.
%%
%%\bibitem{IEEEconf:27}
%%Yong Li, Wenbo Wang, Jia Kong, W.~Hong, X.~Zhang, and M.~Peng,
%%"Power allocation and subcarrier pairing in OFDM-based relaying
%%networks," in \emph{Pro. IEEE ICC 2008}, May 2008, pp.~2602-2606.
%
%%\bibitem{IEEEconf:20}
%%A.~Gersho and R.~M.~Gray, "Vector Quantization and Signal
%%Compression," \emph{Kluwer Academic Publishers}, Boston, 1992.
%%
%%\bibitem{IEEEconf:21}
%%J.~G.~Proakis, Digital Communications, McGraw Hill, ISBN 0-07-
%%118183-0, fourth edition 2003.
%
%%\bibitem{IEEEconf:29}
%%M.~Hajiaghayi, M.~Dong, and B.~Liang,  "Using Limited Feedback in
%%Power Allocation Design for a Two-Hop Relay OFDM System", \emph{Pro.
%%IEEE ICC 2009},  pp.~1-6.
%%
%%\bibitem{IEEEconf:30}
%%Zhang ~G., Zhan ~W., and Qin ~J.,"Power Allocation in
%%Decode-and-Forward Cooperative OFDM Systems Using Perfect and
%%Limited Feedback", \emph{Chinese Journal of Electronics}, Vol.~19,
%%No.~2, Apr.~2010.
%%
%%\bibitem{IEEEconf:31}
%%Luc ~V., Jerome ~L., Onur ~O., and Abdellatif ~Z.,"Rate-Optimized
%%Power Allocation for DF-Relayed OFDM Transmission under Sum and
%%Individual Power Constraints", \emph{EURASIP Journal on wireless
%%comm. and networking}, Vol.~2009.
\end{thebibliography}
\end{document}